# Data Science at Udemy: Agile Experimentation with Algorithms

Larry Wai
Udemy
600 Harrison St.
San Francisco, CA 94107

larry.wai@udemy.com

## ABSTRACT
In this paper, we describe the data science framework at Udemy, which currently supports the recommender and search system. We explain the motivations behind the framework and review the approach, which allows multiple individual data scientists to all become 'full stack', taking control of their own destinies from the exploration and research phase, through algorithm development, experiment setup, and deep experiment analytics. We describe algorithms tested and deployed in 2015, as well as some key insights obtained from experiments leading to the launch of the new recommender system at Udemy. Finally, we outline the current areas of research, which include search, personalization, and algorithmic topic generation.

## CCS Concepts
• Applied computing➡ Education➡ E-learning

## 1. INTRODUCTION
### 1.1 What is "data science"?
"Data science" is an evolving term, which currently has broad usage due to the trending popularity of the concept, especially as a job type [Glassdoor]. In fact, the usage is so broad that the term "data science" has been used interchangeably for "computer science" [Naur], "statistics" [Wu], or "business analytics" in general [Press]. In this paper, we define data science in the more strict sense of modern science, namely the scientific method [Garland]. Such a definition has great practical utility, especially in the consumer Internet industry. The scientific method, i.e. controlled experimentation, as applied in consumer Internet typically involves directly manipulating the user experience, which facilitates establishment of causal relationships in user behavior. Furthermore, controlled experiments allow direct measurements of improvements in essentially any well defined business metric. Last, but not least, the role of the practitioner of "data science", a.k.a. the "data scientist", is much more impactful and well defined when associated with execution of the scientific method itself. In particular for consumer Internet, we believe that the development of algorithms for manipulation of the user experience should naturally be included in the role of the data scientist. We would argue that the data scientist as both a developer of algorithms as well as a practitioner of the scientific method is perhaps the most efficient and agile way to drive consumer Internet business metrics.

### 1.2 What is a data science system?
We define the data science system for any company as the collection of tools and processes needed for execution of the data science process defined above. This includes methods for mining of big data, fast exploration of user behavior data, construction and deployment of machine learned models, setup and extraction of data from controlled experiments, deep analysis of experimental data, and agile methods for update of models and experiment configurations so that the whole data science process can repeat as quickly as possible, ideally only limited by the speed of experimental analysis. Here is a simple diagram depicting the data science process:

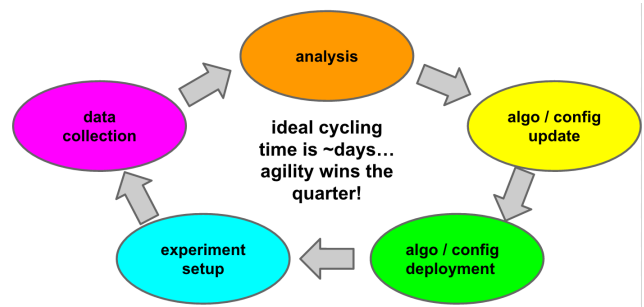

### 1.3 What are the main deficiencies in many data science systems?
We would argue that the main deficiencies in data science systems have to do with limitations of agility. Ideally, the data science process should be limited by collection and analysis of experimental data. Often, even a few days of experimental data collection at a consumer Internet company can already produce enough new information to help inform decision making on next steps. However, in practice it is difficult to reach this ideal limit due to a few factors. First, as in most engineering organizations, work is typically matrixed; that is, the tasks of exploratory analysis, model building, deployment, and experiment analysis are often divided among different people or teams, thus leading to significant delays as the data science process steps through the various stages of the experimental cycle and the results of the previous step need to be communicated to the group or individual responsible for the next step. Furthermore, because the work is matrixed, the workflows used by different teams for different steps, for example experiment analysis vs model building, are often redundant but different systems which make different assumptions about the data and therefore end up creating inconsistencies. Secondly, algorithms are often directly encoded in the production codebase, which creates a high barrier to deployment and update due to the QA processes needed to ensure that production code does not crash the site. Finally, deep experiment analysis is often too slow in execution to facilitate well informed decision making on the timescale of less than a day.

In the spirit of "building a better mousetrap", in the following sections we describe our efforts to "build a better data scientist" by constructing a data science framework at Udemy, which we believe largely addresses the major obstacles for an agile data science process.

## 2. THE UDEMY DATA SCIENCE FRAMEWORK

### 2.1 Summary of Approach

Our basic approach is to create a data science framework, which allows an individual data scientist to execute the entire data science process "with their own two hands." That is, independently perform the exploratory data analysis, build predictive models, deploy predictive models into the production system, setup experiments, analyze the resulting experimental data, make decisions on what to do next, and repeat the whole cycle again within a few days. The advantage of this approach is that the company benefits from a significantly faster data science process, which will in turn increase the rate of improving the algorithms which give lift to key business metrics. Furthermore, individual data scientists will be more incentivized because the lift of business metrics can be more easily attributed to their labor. Finally, because the data scientist has become "full stack", i.e. can execute the entire data science process from end-to-end, a team of such full stack data scientists is more robust, because the individual contributors are no longer specialized / matrixed to particular steps in the data science process and can therefore offer more redundancy among themselves.

### 2.2 Data Mining Workflow

The data mining workflow is designed to be as modular as practical, leveraging a "star schema" data model. At the center of the data model is the largest table, which contains all the courses seen by every visitor to the web site. In order to create a modular workflow, we chose the keys of the central table, a.k.a. the "impression funnel", to be visitor id, course id, and date. All the key measures were included in the impression funnel: click, enrollment, revenue, post-enrollment course video consumption, and post-enrollment Net Promoter Score (NPS). As dimensions were added to the data model the only requirement was that each new dimensions would need to have keys which are subsets of the keys of the impression funnel. For example, static course dimensions would only need course id as key, whereas evolving course dimensions would need course id and date as keys.

The data mining workflow was built in hive / hadoop. One of the key features of the workflow is a configurable framework for auto generating multiple OLAP hypercubes.[1] The modular data model plus configurable hypercubes facilitated the use of the data mining workflow for multiple use cases, including exploratory data analysis, training dataset generation, experiment analysis, as well as trending analysis. By leveraging a single data mining workflow for all of these key use cases, we were able to render unnecessary multiple data mining workflows found in many companies (for example, separate workflows for training dataset generation versus experiment analysis). In addition to eliminating redundant engineering effort, this approach has the added advantage of creating consistency. For example, the same dimensions (in this case features) used for the predictive model training dataset can also be used for experiment analysis; this allowed us to check precisely whether the deployed predictive model was changing the user experience in the expected way.

### 2.3 Predictive Modeling Workflow

The predictive modeling workflow was designed to be as modular as possible, leveraging the PMML [DMG] standard for recording predictive models. We initially used the R packages for creating predictive models (in our case, decision trees) and then saving the resulting models in PMML format. In order to handle the large amounts of training data, we leveraged the OLAP hypercubes from the data mining workflow to aggregate the training data to a size manageable by a simple R setup.[2]

After creating and recording the predictive models in PMML format, the models were subsequently saved into a MySQL database, which serves as a central repository for the predictive models. Java classes, which leverage open source PMML libraries [Ruusmann], were created to facilitate scoring of the predictive models. These java classes were used subsequently in the production workflow, described below, for scoring in real time applications as well in batch mode as a hive UDF. The final step of the predictive modeling workflow is to score holdout data in hive using the PMML based UDF and inspect the residuals between the model prediction and actual business metrics.

### 2.4 Production Workflow

The production workflows have batch as well as real time versions. Both leverage the same PMML based Java classes described above. In the case of batch scoring in hive, we use the same UDFs used for scoring of holdout data in the modeling workflow. The personalized scores on a per course basis are computed for approximately 5 million recent visitors to the site and uploaded to a Redis cache on a nightly basis. The recommendation engine then retrieves the scores on a per visitor request basis and uses them for ranking of courses.

In the case of real time scoring, we created a custom Java based middleware, which handles construction of the feature vectors and PMML based scoring on a per visitor request basis (for both recommendation and search cases).[3] Currently the real time scoring makes use of batch computed features, which are loaded into a cache similar to the batch scoring case. We are currently implementing a real time feature workflow, which will allow streaming of in-session visitor impressions and interactions for in-session re-ranking of courses.

---

1     The methodology for auto generating OLAP hypercubes and their multiple applications will be described in a forthcoming paper [Williams, Han, Wai].

2     Note that in case of aggregated data, we also needed to use weighting in the decision tree configuration in R.

3     The real time scoring engine will be described in a forthcoming paper [Yildirim, Hou, Han, Wai].

## 2.5 Deep Experiment Analytics

Perhaps the most underestimated part of the data science process is the agile production of deep experiment analytics. This is in part due to the relatively trivial process of computing overall experiment metrics, and the much more challenging process of computing binned experiment metrics, even for binning along one dimension..[4]

The basic methodology involves production of a single OLAP hypercube per "numerator" dimension, leveraging the data mining workflow described above. We define a "numerator" dimension as one which does not map onto the keys of the denominator. In our case, we typically use visitorid and datestamp as our denominator for statistics. Any dimensions which map onto visitorid and datestamp would be defined as a "denominator" dimension, and could be dimensions of every hypercube. Conversely, in this case numerator dimensions would not map onto both visitorid and datestamp; for example, the web page context (e.g. featured page, search page, etc.) would not map onto both visitorid and datestamp, since the visitor will in general visit multiple page contexts in a single session.

The cells of each hypercube contain sums of the key measures: impressions, clicks, enrollments, revenue, post-enrollment minutes consumed, and post-enrollment Net Promoter Score(s) (NPS), as well as the associated aggregates needed for T-test statistics. The various hypercubes per numerator dimension are appended together into a single analytics table, which is then ingested into an online Tableau dashboard. The Tableau dashboard then computes the binned average differentials (between test and control variants) and color codes the 95% confidence statistical significance levels. One key feature of our experiment analytics table is that it contains all the historical experiments as well as the current one, so that they can be easily compared by flipping between experiments.

In practice, our Tableau dashboard has a latency of about 1 second between rotations of hypercubes onto different dimensions. A typical analysis session (say 30 minutes) may involve 10 to 100 rotations, including rotations between the various dimensions, filtering, and flipping between historical experiments.

## 3. THE UDEMY GLOBAL TEACHING & LEARNING MARKETPLACE

There are currently many flavors of online education resources, ranging from a more straightforward implementation of existing traditional teachers and courses into an online setting to resources unique to online services. In some sense, the World Wide Web itself is a giant education resource, with search engines being the most obvious tool for finding information to teach oneself through the querying process.

Udemy takes the approach of providing online course creation tools, which helps anybody to create and market a course to a global student body of approximately 10 million. The marketplace is based upon a >=50% revenue share with the instructor, which provides a steady stream of income to the instructor once they have created the course and started collecting course enrollment fees. The marketplace growth is driven by the financial incentive for new course creation on the instructor side, and the corresponding growth on the student side driven by breadth and depth of course selection.

Since many of the top EdTech companies are still privately owned, comparison of online education marketplace performance between companies is difficult. However, Google provides a nice service called "Google Trends", which graphs relative trends of search traffic for different search terms. Using this service, we can estimate the relative search traffic for different EdTech companies over time and get a sense of whether their marketplaces are growing. The following chart shows the trending comparison of Udemy, Udacity, Coursera, and Lynda; as you can see, Udemy does in fact show an exponential growth curve, which is characteristic of a growing marketplace:

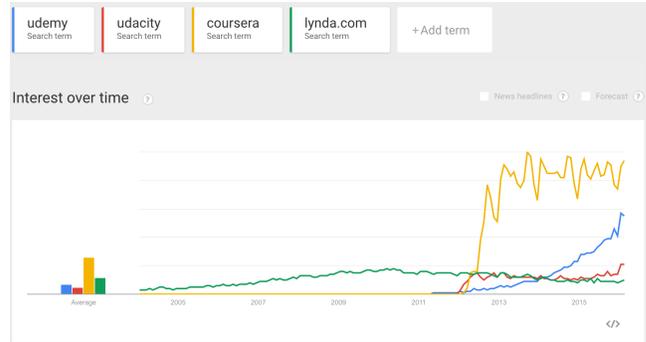

## 4. FEATURED PAGE EXPERIMENTS
### 4.1 Featured Page Ranking Algorithm

The default Udemy web page when arriving at the site (after the first visit) is the "featured" page. This is a general discovery page, which spans across all categories of courses. The featured page is composed of horizontal collections of courses, which we will define as "units." Each unit has a distinct topic or theme. The initial project executed within the Udemy data science framework described above was to improve course recommendations on the featured page. The scope of the featured page experiments included improving the ranking of the existing units, as well as improving the ranking of the courses within the units. Our initial experiments did not result in the introduction of new units, although this is planned for the next generation of experiments.

The baseline for the experiments was a rule based ordering of the units, and a randomized ordering of courses within the units, which would be re-ranked whenever the visitor would re-visit the featured page. Each unit contains 24 courses, of which 4 are shown by default and up to 8 more shown by clicking on an expansion button. The units are defined either by collaborative filter based on some visitor action, or by some other heuristic: "because you searched for X", "because you enrolled in X", "students who viewed X also viewed", "new and noteworthy", "students are now viewing", "bestsellers in Y".

Our basic strategy for ranking is to compute a predicted personalized score per course, and use that score for ranking of courses within the units as well as ranking of the units on the page. In order to rank the units, we computed a unit score, which is the sum of the personalized course scores for the top 4 courses in each unit. After finding the top scoring unit, we removed the top 4 courses from that top scoring unit for the purposes of recomputing the unit scores for the remaining units. By repeating

---

[4] We will describe / teach in detail the methodology for agile computation of binned experiment metrics in a forthcoming Udemy course.

this process, we were able to generate an optimized de-duped set of courses for the page on the initial view (since a typical browser size shows by default the top 4 courses in each unit). The following diagram depicts the ranking algorithm described above:

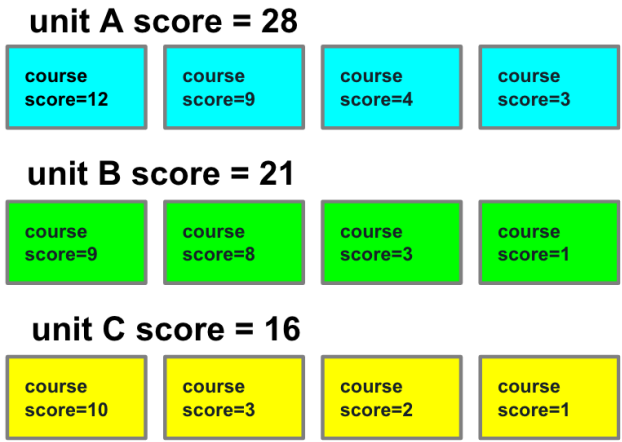

## 4.2 Factorized Predictive Models

Data mining was performed on two basic datasets, which we call the "impression funnel" and the "enrollment funnel." The impression funnel has as its basic unit unique course impressions per visitor per datestamp. The enrollment funnel has as it's basic unit course enrollments per user. The impression funnel was used to build a predictive model for the enrollment rate per thousand impressions (EPMI), and the enrollment funnel was used to build predictive models for consumption per enrollment (CPE) and NPS per enrollment (NPE). The basic idea for creating these two kinds of predictive models is that we can use them as multiplicative factors in the overall score used for ranking, and thus "mix and match" to create more testable variants with fewer models. In general, the scoring function can be defined as follows:

$$\text{score} = \text{EPMI} \times P^{\alpha} \times \text{CPE}^{\beta} \times \text{NPE}^{\gamma}$$

where P is the price of the course and $\alpha$, $\beta$, $\gamma$ are exponential parameters which can be configured according to the desired targeting business metric. For example, if we want to target enrollments only, then we would set $\alpha=0$, $\beta=0$, $\gamma=0$; similarly, if we want to target video consumption, then we would set $\alpha=0$, $\beta=1$, $\gamma=0$. We can even target blended versions of the business metrics by setting the parameters to some fractional value between 0 and 1.

Our data mining approach was to use regression trees, with some trimming to avoid overfitting. The reason for using regression trees rather than an ensemble approach like random forest was to reduce latency in the real time scoring scenario. We used two basic kinds of modeling features, historical course averages (in the trailing 91 days) as well as personalized features based upon per course and per subcategory clicks (positive feedback) or lack of clicks on seen course impressions (negative feedback). The personalized positive / negative feedback features will be described in the following section.

## 4.3 Explore and Exploit Algorithm

### 4.3.1 Course interest feature
The most basic element of our explore / exploit algorithm is what we will define as the visitor "course interest feature." This feature measures the interest a visitor has expressed in a particular course; the interest can in general be defined in terms of clicks on the course impression normalized by the number of impressions, weighted by the recency of the impressions. In the case of our initial experiments, we have chosen perhaps the simplest definition of course interest: considering only the last impression of the course, did the visitor click on it (positive interest) or not (negative interest). If the course has not yet been seen, then we consider the course interest to have a null value. As a naive Bayes classifier for EPMI, the ratios of negative interest : null : positive interest are approximately 0.8 : 1.0 : 3.1.

### 4.3.2 Subcategory interest feature
In general, we can extend positive / negative interest in individual courses to imply interest (or lack of) in as yet unseen courses similar to courses which have already been seen, where the similarity can be defined along arbitrary dimensions (e.g. topic, course difficulty, teaching presentation style, etc.). Mathematically, we can express the interest in a collection of similar courses K in terms of our course interest feature:

$$\text{interest in K} = \Sigma \, [f(\text{course interest}) \times g(\text{course})] \, / \, \Sigma \, g(\text{course})$$

where the sum is over courses in collection K seen by the visitor, f is a function of the course interest, and g is course dependent weight. In our initial experiments, we considered collections of courses in subcategories (i.e. K = subcategory). We defined a subcategory interest feature to be the % of courses seen by a visitor in a subcategory in the trailing 91 days which were clicked by the visitor. In the null case where the visitor has not seen any courses in the subcategory yet, we assigned a default prior of 5%.

### 4.3.3 Explore and exploit optimization
The following chart shows how our explore and exploit algorithm works. The basic idea is to log all the courses seen by visitors, as well as any interactions with those seen courses, like clicks. The negative feedback is as important as the positive feedback for input into the algorithm; this can be contrasted with typical collaborative filters, which only use the positive feedback as input. As with collaborative filters, the clicked inventory (or similar inventory) is exploited by boosting the score (and ranking). However, in our algorithm we also exploit the lack of clicks on inventory (or similar inventory) by penalizing the score (and ranking). By putting some suppression on seen inventory with negative feedback, this opens up space for unseen inventory to be explored. This method naturally incorporates the need for "freshness" in a balanced and systematic way, i.e. we do not compromise the inventory with positive feedback, while promoting freshness.

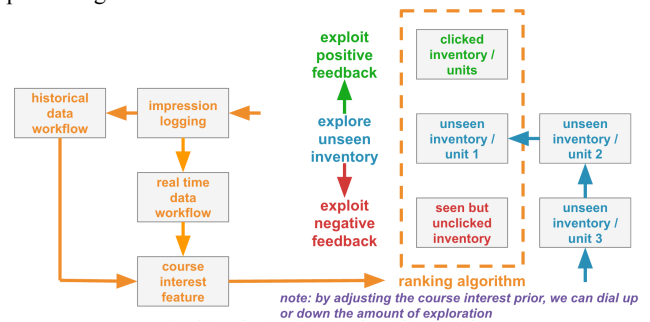

In our initial experiments, we used the course interest feature and subcategory interest feature as features in the regression tree modeling. We are planning new experiments to use these features as naive Bayes classifiers, and also test "boosting" of the priors, which will allow us to experiment with dialing up or down of the amount of exploration, and thus optimize the explore / exploit algorithm against business metrics. Another class of experiments planned are to use human curation as input for either course or subcategory priors. The idea is to leverage human curation to help with the cold start problem; for example, if our content managers believe that a new course will have a high enrollment rate, we can set the prior for the course interest feature to be a higher than average value.

## 4.4 Experiment Results

The featured page experiments were run over a 3 month period, with roughly 20 configuration / model updates, before being launched to 90% of English language visitors in November of 2015. The agility of the data science framework was crucial for success. We were able to collect experimental data on each configuration or model update, quickly perform deep analysis and within a few days determine potential issues, which needed to be fixed, and deploy the new configurations or models into the production system. If, for example, it would have taken 2-3 weeks for deep analysis plus configuration / model update in the production system (typical at many companies), then it would have required a full year of experimentation to arrive at the same launch candidate.

A final experiment was run for 5 weeks before making the new recommender system the baseline for 100% of English language visitors on the featured page. The final experiment results indicate an increase of +2% in visitor traffic, an overall increase of +7% for new enrollment video consumption, an overall increase of +3.4% for new enrollment NPS, and an overall increase of +7.5% in revenue per session. The lift was larger for the featured page itself: +17% for new enrollment video consumption, +36% for new enrollment NPS, and +15% in revenue per session.

## 4.5 Key Insights

### 4.5.1 Quality generates more revenue

We ran several variants with different scoring function target metrics: enrollment, revenue, consumption, and NPS. What we found is that impression normalized enrollment weighted NPS (EPMI x NPE) was the targeting metric, which generated the best overall lift in business metrics, including revenue. In other words, the revenue targeting variant actually generated less revenue than the NPS targeting variant.

Our interpretation is that session level effects may play a strong role here. If the variant ranks too many high priced courses towards the top of the page, then the visitor may be less enthusiastic about scrolling down lower or otherwise continuing their session on the site. By targeting course quality (i.e. NPS), the visitors have a better experience earlier in their session, which leads to longer sessions and ultimately higher per session conversions.

### 4.5.2 Strong assisted revenue effects

A striking aspect of the experiments was that by improving the recommendation algorithm on the featured page, we saw a lift of engagement and business metrics across the entire site. In fact, only 25% of the total revenue lift was on the featured page itself; most of the absolute revenue gain was from pages unaffected by our algorithm change. Perhaps most surprising, was the observation that even direct landings on courses through e-mail clicks had a +5% lift of revenue. Our working hypothesis is that since the featured page is the default page on the web site, by improving the user experience there, visitors were more motivated about Udemy in general, to the point that they even paid more attention to their Udemy marketing e-mails. We are currently working on a deeper path analytics framework, which should help us to more clearly track the sequence of actions of visitors at Udemy (whether in search, discovery, e-mail, or consumption) and thus demonstrate causality between various activities. The following chart shows the revenue lifts on the various pages / properties at Udemy due to introduction of the improved recommendation algorithm on the featured page.

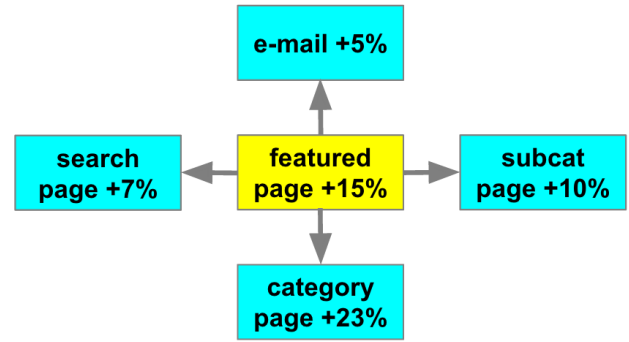

### 4.5.3 More can be less

One interesting observation is that by eliminating low conversion units on the featured page, we reduced the number of impressions on the featured page, but conversions on the featured page went up as well as engagement and conversions across the site. Our interpretation was that this was a good example of "more is less", i.e. people were previously wasting more time on looking at the courses in the low conversion units and spending less time on the higher converting units on the featured page as well as exploring the rest of the site.

### 4.5.4 Free courses have a context dependent role

Free courses play an interesting role at Udemy. In our experiments, we noticed that decreasing the number of free courses seen on the featured page increased revenue on the featured page, but also reduced revenue on the rest of the web site so that the net revenue was actually neutral to negative. Only by optimizing the mix of free and paid courses on the featured page were we able to obtain an overall net revenue increase across the site while maintaining revenue lift on the featured page. (The optimum is about 25% free courses seen in a user session.) Our interpretation of this is that free courses have the effect of increasing user engagement so that visitors spend more time exploring the site and end up generating more revenue overall. This effect appears to be context dependent; when we tried increasing the number of free courses on the post-enrollment page, the overall revenue went down. Our interpretation is that showing free courses earlier in the user session (like on the featured page) has a bigger impact on increasing the user engagement and subsequent conversions, as compared to showing free courses later in the user session (like on the post-enrollment page) which may only reduce the revenue at that late stage in the session.

## 5. NEXT STEPS

After deploying the new recommendation algorithm to the featured page on the web site, we are planning to continue deploying the algorithm to other pages on the web site, e-mail, and mobile. We are also experimenting with the search algorithm using the same data science framework.[5] We are currently researching how to improve the recommendation predictive models to include more personalization.[6] Also, we are researching an algorithmic method for generating course topics and how to introduce the algorithmically generated topics into new units on the web site.[7]

## 6. ACKNOWLEDGMENTS

Four related teams have contributed to the data science framework at Udemy and subsequent experiments: Data Science, Data Infrastructure, Recommendation, and Search. The Data Science team includes: Larry Wai (principal), Imeh Williams, Jian Yang, Beliz Gokkaya. The contributors from the Data Infrastructure team include: Keeyong Han (principal architect), James Hou, Mars Williams, Sungju Jin. The recommendation team includes: Gulsen Yildirim (engineering manager), Cetin Cavdar, Cagatay Calli, Melda Dadandi, Erol Aran. The search team includes: Okan Kahraman (engineering manager), Ahmet Korkmaz, Ozgen Muzac, Heval Azizoglu, Ibrahim Tasyurt.

---

[5] The methodology for search ranking will be described in a forthcoming paper [Williams, et.al.].

[6] The methodology for personalization will be described in a forthcoming paper [Yang, et.al.].

[7] The methodology for algorithmic topic generation will be described in a forthcoming paper [Gokkaya, et.al.].